\documentclass[journal,comsoc,12pt,onecolumn,draftclsnofoot]{IEEEtran}

\usepackage[T1]{fontenc}
\usepackage[utf8]{inputenc}
\usepackage{amsmath}
\usepackage[cmintegrals]{newtxmath}
\usepackage{color}
\usepackage{xcolor}
\usepackage{cite}
\usepackage{etoolbox}
\makeatletter

\usepackage{enumitem}
\usepackage{url}

\usepackage[binary-units = true, range-phrase=--, per-mode=symbol]{siunitx}
\DeclareSIUnit{\decibelm}{dBm}
\DeclareSIUnit{\Joule}{Joule}
\DeclareSIUnit{\Sample}{S}

\ifCLASSINFOpdf
   \usepackage[pdftex]{graphicx}
   \graphicspath{{Figures/}}
\else
\fi

\usepackage{changes}

\makeatletter
\newcommand\RCC[1]{%
  \setkeys{Changes@definechangesauthor}{color=#1}%
  \expandafter%
  \let\csname Changes@AuthorColor\endcsname=\Changes@definechangesauthor@color%
  \colorlet{Changes@Color}{\@nameuse{Changes@AuthorColor}}%
}
\makeatother

\begin{document}

\title{Autonomous Interference Mapping for \\Industrial IoT Networks over Unlicensed Bands}

  \author{Simone~Grimaldi,
          Aamir~Mahmood, 
          Syed~Ali~Hassan,
					Gerhard~Petrus~Hancke,
          and~Mikael Gidlund
  \thanks{S. Grimaldi, A. Mahmood and M. Gidlund are with the Department of Infomration Systems and Technology, Mid Sweden University, Sweden. e-mail: simone.grimaldi@miun.se}
  \thanks{S. A. Hassan is with the School of Electrical Engineering and Computer Science (SEECS), National University of Sciences and Technology (NUST), Islamabad, Pakistan.}
	\thanks{G. P. Hancke is with the  Department of Computer Science, City University of Hong Kong.}%
  }

\maketitle

\begin{abstract}
The limited coexistence capabilities of current Internet-of-things (IoT) wireless standards produce inefficient spectrum utilization and mutual performance impairment. The entity of the issue escalates in industrial IoT (IIoT) applications, which instead have stringent quality-of-service requirements and exhibit very-low error tolerance.
The constant growth of wireless applications over unlicensed bands mandates then the adoption of dynamic spectrum access techniques, which can greatly benefit from interference mapping over multiple dimensions of the radio space.
In this article, the authors analyze the critical role of real-time interference detection and classification mechanisms that rely on IIoT devices only, without the added complexity of specialized hardware. The trade-offs between classification performance and feasibility are analyzed in connection with the implementation on low-complexity IIoT devices.
Moreover, the authors explain how to use such mechanisms for enabling IIoT networks to construct and maintain multidimensional interference maps at run-time in an autonomous fashion. Lastly, the authors give an overview of the opportunities and challenges of using interference maps to enhance the performance of IIoT networks under interference.

\end{abstract}


\IEEEpeerreviewmaketitle

\section{Introduction}

\IEEEPARstart{U}{nlicensed} bands are playing a pivotal role in the diffusion of the Internet-of-things (IoT) paradigm in several domains. Smart-homes, building automation~\cite{smart_home,building_automation}, and industrial monitoring and control~\cite{car_industry} are notable applications that significantly benefit from short-range, low-power communication. While the upcoming 5G wireless technology promises to fulfill many requirements of such systems~\cite{5g_survey}, it overlooks two of the essential advantages of unlicensed spectrum: worldwide availability of frequencies and license-free network deployment~\cite{5G_automation}. Other than a significant cost-containment, short-range transmissions over unlicensed band ensure complete control over the data-flow (e.g., process-control data) and network infrastructure.
Nonetheless, an IIoT network operating over unlicensed bands is exposed to the risks of non-exclusive spectrum usage, i.e., interference, as co-located in-band wireless systems interact destructively with each other.

In the popular \SI{2.4}{\giga\hertz} unlicensed band, for instance, several technologies (i.e., IEEE 802.11-based WLANs, IEEE 802.15.1, and IEEE 802.15.4 networks) compete for spectrum access without an inter-technology coordination mechanism, which causes mutual interference.
As the generated interference level shows complex dynamics in time, space and frequency, the network-wise interaction can become rather unpredictable~\cite{survey_WSN_interf}, posing severe threats to communication quality.
This scenario discourages the implementation of mission-critical IIoT applications (such as industrial monitoring and control) over unlicensed bands unless proper mechanisms for inter-technology coordination or interference avoidance are adopted. 
Since inter-technology coordination and spectrum collaboration are still at their early development stages~\cite{darpa}, the importance of spectrum sensing and interference mitigation becomes crucial.

Spectrum sensing aims at quantifying the spectrum usage statistics or at identifying the interfering signals across a frequency band at a given location. 
The aggregation of distributed spectrum measurements provides a representation (or map) of interference over a multidimensional radio space and gives insights on how the interference vary within the location of interest~\cite{IMap_80211}, e.g., the industrial plant. The availability of such maps enables opportunistic spectrum access approaches, such as interference-aware channel selection~\cite{blacklist} or interference-aware routing~\cite{IMap_Routing}.

The existing literature has investigated spectrum monitoring solutions for interference maps~\cite{Hoyhtya_InterferenceMaps} and electrosmog monitoring using big data~\cite{Rajendran_Electrosense}.
However, to the best knowledge of the authors, there is a lack of interference mapping solutions expressly designed for IIoT systems with low-complexity hardware.
A reason for that is the complexity of the measurement and analysis process, which collides with the constrained sensing and computational capabilities of network devices.
As an additional challenge, IIoT devices can only rely on energy samples with limited time and frequency resolution.
Therefore, many works in the literature focus on increasing the energy-sampling rate or sensing for a longer time.
Unfortunately, this has substantial drawbacks of pushing the requirements on storage size and potentially disrupting the routine operation of IIoT networks.
Also, as the sensing bandwidth increases~\cite{sensing_bandwidth}, the processing requirements easily fall out of the reach of IIoT platforms. 

This article analyzes the challenges and opportunities of wireless-technology-specific interference maps in the presence of several co-located wireless networking technologies.
The key role of real-time RF interference analysis using resource constrained IIoT radio devices is discussed and the potential of lightweight signal features and machine-learning classification is examined. 
Finally, the paper analyzes a real use case that encompasses WLAN and Bluetooth interference mapping with IEEE 802.15.4 devices and discusses the use of the resulting fine-grained RF maps for RF coexistence enhancement over different time horizons.


\section{Coexistence in Unlicensed Bands}
\label{sec:background}
\subsection{Cross-Technology Interference}
WirelessHART~\cite{WHART} and ISA 100.11a~\cite{ISA100} are the two leading wireless standards used in the process automation domain~\cite{IWSN_R3}.
Both standards are based on the IEEE 802.15.4~\cite{802154} physical layer, which operates in the \SI{2.4}{\giga\hertz} unlicensed band.
The IEEE 802.15.4 standard is also the basis for ZigBee and 6LoWPAN solutions, which are largely employed in building automation applications~\cite{build_autom_survey}.

The \SI{2.4}{\giga\hertz} unlicensed band provides \SI{85}{\mega\hertz}-wide spectrum available worldwide, despite minor regional variations to its extension.
The availability of free spectrum with a relatively wide bandwidth has stimulated the massive usage of various wireless standards, such as IEEE 802.11 and IEEE 802.15.1, Bluetooth Low Energy (BLE), and the aforementioned IEEE 802.15.4, which constitute the basis for several modern IoT applications. 
MulteFire~\cite{multe_fire_R4}, which extends the use of LTE in unlicensed (\SI{2.4}{\giga\hertz}/\SI{5}{\giga\hertz}) and shared bands, is another recent candidate technology for communication over unlicensed spectrum.
MulteFire Release 1.1 specifically targets enhanced machine-type communication at \SI{2.4}{\giga\hertz}.

Due to the absence of a shared mechanism for spectrum access coordination, wireless systems based on different standards show a destructive interaction when coexisting in the same environment.
The interaction manifests in the form of i) packet drops, meaning that multiple radio signals overlap in time-frequency at the receiver, and the decoded information is corrupted, and ii) delayed transmissions, due to clear-channel-assessment (CCA) and collision-avoidance mechanisms. Such effects, classified as \textit{collision loss} and \textit{inhibition loss}~\cite{adaptive_CCA}, respectively, can drastically reduce the performance of the affected networks~\cite{Surviving_Wifi_R4,petrova_wifi}.
The performance impairment poses serious concerns not only for applications with strict reliability or latency requirements (i.e., industrial applications) but also for non-critical IoT applications due to QoS reduction and increased energy consumption.


For the reasons above, cross-technology interference is widely acknowledged as the principal cause of interference in the \SI{2.4}{\giga\hertz} band.
Radio-frequency (RF) interference from electric equipment and consumer electronic devices represents indeed a relatively secondary concern, given the brevity and low-energy of RF noise spikes~\cite{noise_WLAN_band,rapaport} at \SI{2.4}{\giga\hertz}.
The only exceptions are some digital cordless phones and baby monitors~\cite{TIMO_R4}, and the high-power emissions of microwave ovens~\cite{rapaport}, which nevertheless find limited application in industrial scenarios.

\begin{figure}[!t]
\centering
\includegraphics[width=0.75\textwidth,clip, trim=0cm 0cm 0cm 0cm]{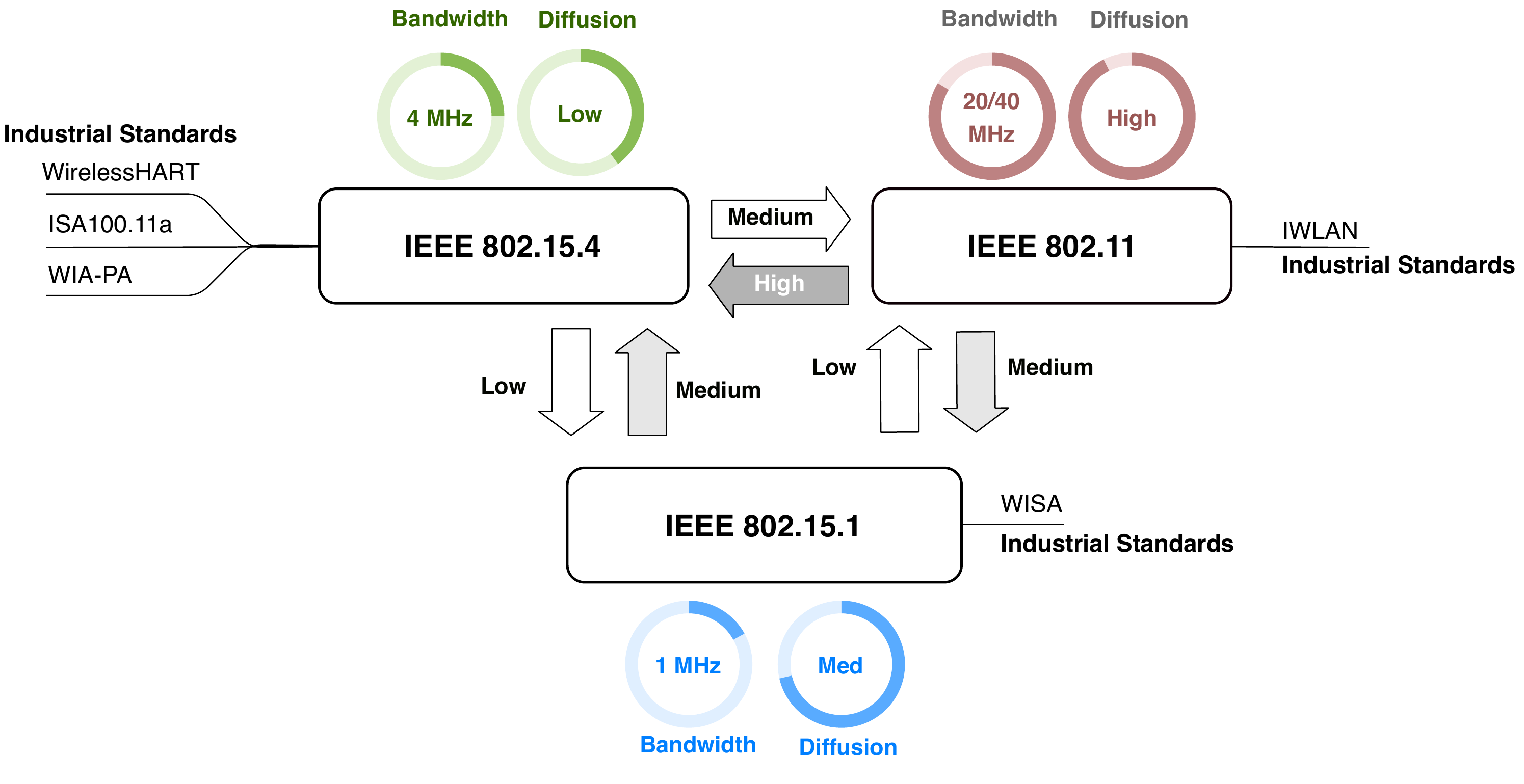}
\caption{An overview of IoT wireless technologies with a focus on cross-technology interference impact. WIA-PA: Wireless networks for Industrial Automation - Process Automation, IWLAN: Industrial Wireless LAN, WISA: Wireless Interface for Sensors and Actuators.}
\label{FIG:IOT_technologies}
\end{figure}

Fig.~\ref{FIG:IOT_technologies} highlights the characteristics of IoT radio standards, which show significant differences with respect to the spectrum occupancy and transmission power.
This, in turn, reflects into a strongly asymmetric effects of cross-technology interference~\cite{WLAN_vs_WSN}.
In particular, the IEEE 802.11 transmissions are a severe issue to IEEE 802.15.4 and IEEE 802.15.1  systems~\cite{mutual_int} due to high transmission power and broad spectrum. The effects of co-located IEEE 802.11 WLANs on the performance of IEEE 802.15.4 networks can become particularly harsh as the attenuation between the interfering and victim devices decreases~\cite{yuan_coex_2013}, or their frequency separation margin reduces (i.e., there is consistent spectral overlap)~\cite{WLAN_vs_WSN}. 
The employed CCA mechanism plays also a major role in shaping the impact of interference by altering the balance between the impact of the aforementioned inhibition and collision losses. 
The related literature has indeed shown that CCA can lead to quite different interference-induced performance impairment~\cite{Duquennoy_R4}, and that employing adaptive CCA schemes (i.e., the energy threshold used for CCA is dynamic) can have beneficial effects on the throughput of IEEE 802.15.4 links under IEEE 802.11 interference~\cite{adaptive_CCA}.

\subsection{Interference Awareness in IIoT}
Identifying the nature and characteristics (e.g., packet length, bandwidth, and power) of the network causing the interference is fundamental to predict the performance impact on the victim network~\cite{lenght_vs_per}.
Nevertheless, current IIoT systems only implement non-specialized interference mitigation techniques, such as periodic and a pseudo-random transmission channel switching (channel hopping~\cite{ISA100}) or deferring the transmission based on bare threshold-based energy detection.
The threshold-based energy detection approach is still followed in the new releases of the IEEE 802.11 standard, e.g., IEEE 802.11ax, despite its drawback of negligible performance improvements in dense deployments~\cite{wifi_cs}.
The lack of appropriate coexistence mechanisms in IIoT standards has stimulated several approaches in the literature under the cognitive radio paradigm. 
The main idea of cognitive radio is to profile and leverage the evolution of radio-resource usage in a hyperspace composed of multiple dimensions, i.e., space, frequency, time, and  possibly others~\cite{Cognitive_Industrial}. The information is generally acquired through spectrum sensing, wherein the radio devices autonomously scan the available spectrum to detect and profile the interference (i.e., the other radio-networks) in time, frequency or other domains.
Nevertheless, a relevant problem of implementing such approaches in real IIoT networks is that radio nodes have typically limited hardware capabilities, both from the sensing and the processing point of view, inhibiting the use of more advanced signal processing techniques, such as interference cancellation~\cite{int_canc_R4}.
A second challenge is the harmonization of the scanning and processing time with the network schedule. In fact, as discussed in the next Section, the interference identification tasks can be rather time-consuming and demand optimized sensing and classification methods.

\section{Measuring Interference with Resource-constrained Devices}
\label{sec:ITI_Motivation}
\subsection{Feasibility of Energy Sampling}

The availability of an efficient spectrum scanning method on constrained hardware is fundamental requirement for interference mapping in IIoT. 
Common spectrum scanning relies on energy-sampling, which means that the sensing device is used as a microwave radiometer to sample and store the energy level at its radio front-end. The samples-trace is then analyzed to reveal specific characteristics of the acquired signal.
Unfortunately, the substantial complexity reduction due to the spared additional hardware is paid with reduced sensing performance~(see \cite{Detecting_non_wifi,Simone_RT_IDI}), as energy-sampling is a barely auxiliary feature for low-cost radio devices.
Its accuracy is indeed far inferior to specialized hardware, expressly conceived for spectrum measurements (e.g., spectrum analyzers), due to both radio and signal-processing limitations.
Also, the sensing performance can be substantially diverse among different hardware platforms, as the radiometric characteristics of the devices are subject to the requirements of the employed standard.

Table~\ref{TAB:parameters} describes the salient characteristic of an energy-sampling system, including reference parameters of an IEEE 802.15.4-based platform.
\bgroup
\begin{table*}[ht]
\caption{General Parameters for Energy Sensing With IoT Hardware, and Practical Example from a Low-cost IEEE 802.15.4 Platform. (*) Indicates Solution-Specific Values~\cite{Simone_RT_IDI}.}
\centering
\begin{tabular}{p{0.18\linewidth} p{0.57\linewidth} p{0.15\linewidth}}
 \hline
 \textbf{Parameter} & \textbf{Description} & \textbf{Example Value} \\ \hline 
  \textbf{Accuracy} & Measurement error due to discrepancies among devices. It can be reduced with calibration~\cite{RSSI_cal}. & \SI{+-6}{\decibel}\\
 \textbf{Dynamic Range} & Measurable power range at the radio front-end.  & \SIrange{-100}{0}{\decibelm}\\
\textbf{Frequency Range} & Observable frequency range.  & \SIrange{2400}{2485}{\mega\hertz} \\
 \textbf{Frequency Granularity} & Minimum allowed frequency-step for programming the sensing device.&  \SI{1}{\mega\hertz}\\ 
 \textbf{Sampling Rate}& Rate of energy-sampling, defined by the polling-rate of the RSSI register. & \SIrange{10}{30}{\kilo\Sample/\second}\\
\textbf{Resolution Bandwidth (RBW)} & Capacity to resolve unmodulated carriers spaced RBW-\si{\mega\hertz}. It is constrained by the channel-filter of the device. & \SIrange{1.5}{4}{\mega\hertz} \\
 \textbf{Sample Width} & The bits used for representing each energy-sample. & \SI{8}{\bit}\\
 \textbf{Observation Time (*)} & Time for completing sensing and identification tasks on a single PHY channel.  &\SIrange{30}{50}{\milli\second} \\
 \textbf{Dwell Time (*)} & Time required for a complete spectrum scan.  & \SIrange{600}{800}{\milli\second}\\ 
 \hline
\end{tabular}
 \label{TAB:parameters}
\end{table*}
\egroup
Practically, energy sampling with low-complexity platforms impairs signal detection accuracy due to several reasons. First, the operating frequency of the onboard microcontroller caps the energy sampling-rate, which can be orders of magnitude below the Nyquist-rate, thus compromising signal analysis based on the discrete Fourier transform (DFT).
Moreover, the energy samples are commonly available only in the form of received signal strength indicator (RSSI). 
As the RSSI is typically a heavily-filtered~\cite{802154} real-valued version of raw I/Q samples, it undergoes a significant loss of information on the envelope of the acquired signals. 
Additionally, a limited sample-width and uncalibrated radio front-ends further complicate the extraction of time- and frequency-domain signal-features.

\subsection{The Identification of Interfering Signals}
Interference-technology identification (ITI) is the inference of the technology sources based on the observation and analysis of some manifestation of the interference. The benefit of ITI is that it adds an extra dimension (i.e., interference technology) to the radio space hyperplane, significantly enhancing the cognitive radio approach. Indeed, the identification of coexisting technology can lead to entirely different interference-mitigation strategies, as the different wireless standards have different spectral footprints and a substantially asymmetric impact on each other (see Fig.~\ref{FIG:IOT_technologies}).

Given the relatively limited number of IoT radio technologies, a trivial solution to interference mapping is the deployment of devices equipped with multiple-radio front-ends.
Such a system could natively collect packet-based information on the transmissions of each interfering radio system and map the different coexisting networks. 
Unfortunately, the solution has a notable pitfall of added complexity and costs both on the hardware and the network side. 

A more suitable solution for IIoT is ITI with low-complexity devices only, as it prevents the cost and complexity of deploying dedicated sensing hardware. 
On the other hand, such a system has to deal with the limitations discussed earlier in Section~\ref{sec:ITI_Motivation} and requires significant engineering efforts for the signal-processing and signal-classification steps. 
Finally, a crucial aspect is the selection of a suitable classification method, since it impacts both the ITI accuracy and the practical feasibility with hardware-constrained devices.
Section~\ref{sec:RT_ITI} examines the latter aspect further and discusses how to achieve high performance on ITI even with classifiers of limited-complexity.

\subsection{Generating Interference Maps}
\begin{figure*}[ht]
\centering
\includegraphics[width=0.8\linewidth]{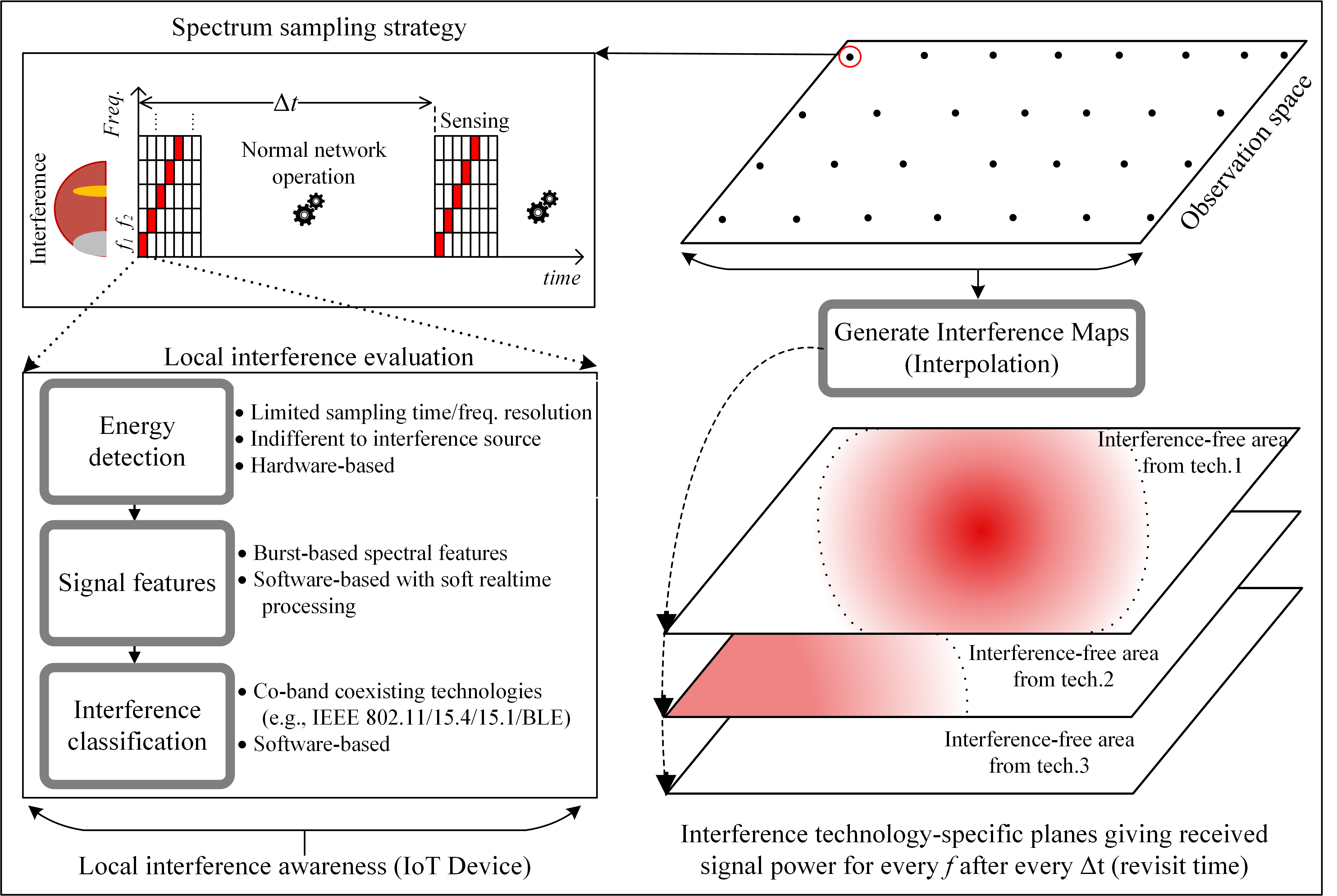}
\caption{From spectrum-sampling to the generation of global interference maps.}
\label{FIG:block}
\end{figure*}

An interference map is a multidimensional representation of specific physical characteristics (e.g., power, busy-time) of RF signals in time-frequency-space or some composite radio space. IIoT network nodes with onboard ITI capabilities can natively generate local time, frequency, and technology characterization of the interference. Later, the function of global map aggregation, which augments the map-dimensionality by adding the space-variable, is operated by a central device, such as a network-manager. Fig.~\ref{FIG:block} gives a graphical representation of the process.

Since the position of interfering devices is not generally known, the preferable distribution of sensing devices is a regular grid-layout, with the grid-spacing chosen according to the detection range of interfering signals~\cite{Hoyhtya_InterferenceMaps}. The trade-off of utilizing IoT devices alone is that the network topology is usually fixed and far from regular, due to network deployment constraints, which \textit{de facto} limits the detection of interfering devices. 
Similarly, the availability of tightly synchronized (i.e., in the order of magnitude of microseconds or lower) distributed measurements can enable mapping of temporal dynamics at the level of single interference bursts. Unfortunately, mapping such temporal dynamics in IIoT networks is rarely an option since maintaining tight synchronization introduces channel and energy overhead. In WirelessHART, for instance, time synchronization is required for time-slotted channel hopping; however, synchronization accuracy is relaxed to a few hundreds of microseconds~\cite{802154_sync} by designing timeslots of \SI{10}{\milli\second} with large guard times.


The process of constructing interference maps has to deal with interference measurements that are inherently discrete in time, frequency, and space. In particular, the space domain can present a relevant sparsity due to the irregular distribution of sensing devices in the deployment area. 
When the interference level among the sampling station is relevant to the cognitive radio application, interpolation and extrapolation techniques can be employed~\cite{Hoyhtya_InterferenceMaps}. The selection of the specific technique depends on the regularity of the distribution of sample points, the particular interference feature of interest, and the constraints on computational complexity.
In particular, such techniques can be \textit{exact} or \textit{non-exact}, depending on if the predicted values at the measurement points coincide with the measured value, and \textit{deterministic} or \textit{stochastic} if the uncertainty distribution of the measured and interpolated values are taken into account~\cite{Hoyhtya_InterferenceMaps}.

%
\section{The Role of Machine Learning in Real-time Identification} 
\label{sec:RT_ITI}
The wide gap between the processing resources of IoT devices and dedicated hardware has traditionally split the research work on ITI into two branches: a) methods operating with high-resolution IQ data and machine learning (ML)-based classifiers, b) methods with RSSI observations from IoT devices and lightweight algorithmic identification methods. Additionally, the use of simpler ML methods based on supervised learning---e.g., support vector machines (SVMs)---that employ data from COTS devices is also examined in some works~\cite{Detecting_non_wifi}.

\subsection{The Importance of Signal Features}
The extraction of appropriate signal features plays a pivotal role in enabling the performance boost of ML methods, and it is particularly critical when operating with heavily-filtered datasets.
For example, IEEE 802.15.4 devices provide RSSI signal traces with bandwidth in the order of only a few \si{\kilo\hertz}, which leads to a poor representation of the signal envelope.
The effect inhibits the classification accuracy of single signal bursts. A recent work ~\cite{Simone_RT_IDI} shows that such information can be partially recovered, as IEEE 802.15.4-compliant hardware is capable of frequency sweeping within the typical duration of interference bursts.
This technique can be used for extracting spectral features (SFs), which contain partial information on the bandwidth of the interfering signals.
The features are particularly relevant for classification, as the different IoT wireless technologies have different spectral footprints.
The extraction of SFs has two significant positive effects as, i) the interference source can be identified based on single signal bursts even with limited sampling frequency and, ii) the classification accuracy increases by \SIrange{10}{25}{\percent} when SFs are included.

\subsection{Lightweight Supervised Learning}
The interest in ML for signal processing and classification is on a rising trend~\cite{ML_survey}.
However, ML methods have so far found limited applicability in run-time ITI approaches due to complexity barriers~\cite{Hermans}. A fairly common solution for constrained platforms is to use heuristic algorithms~\cite{Iver_detecting_and, King_CCA_Variations}. The use of such algorithms may lead to insufficient performance when the target classes are not linearly separable in the selected feature-space~\cite{SONIC_R1}. Besides, as several  supervised-learning methods are available, it is interesting to investigate the existence of a trade-off between complexity and classification performance.

\begin{figure}[!t]
\centering
\includegraphics[width=0.85\textwidth,clip, trim=0cm 0cm 0cm 0cm]{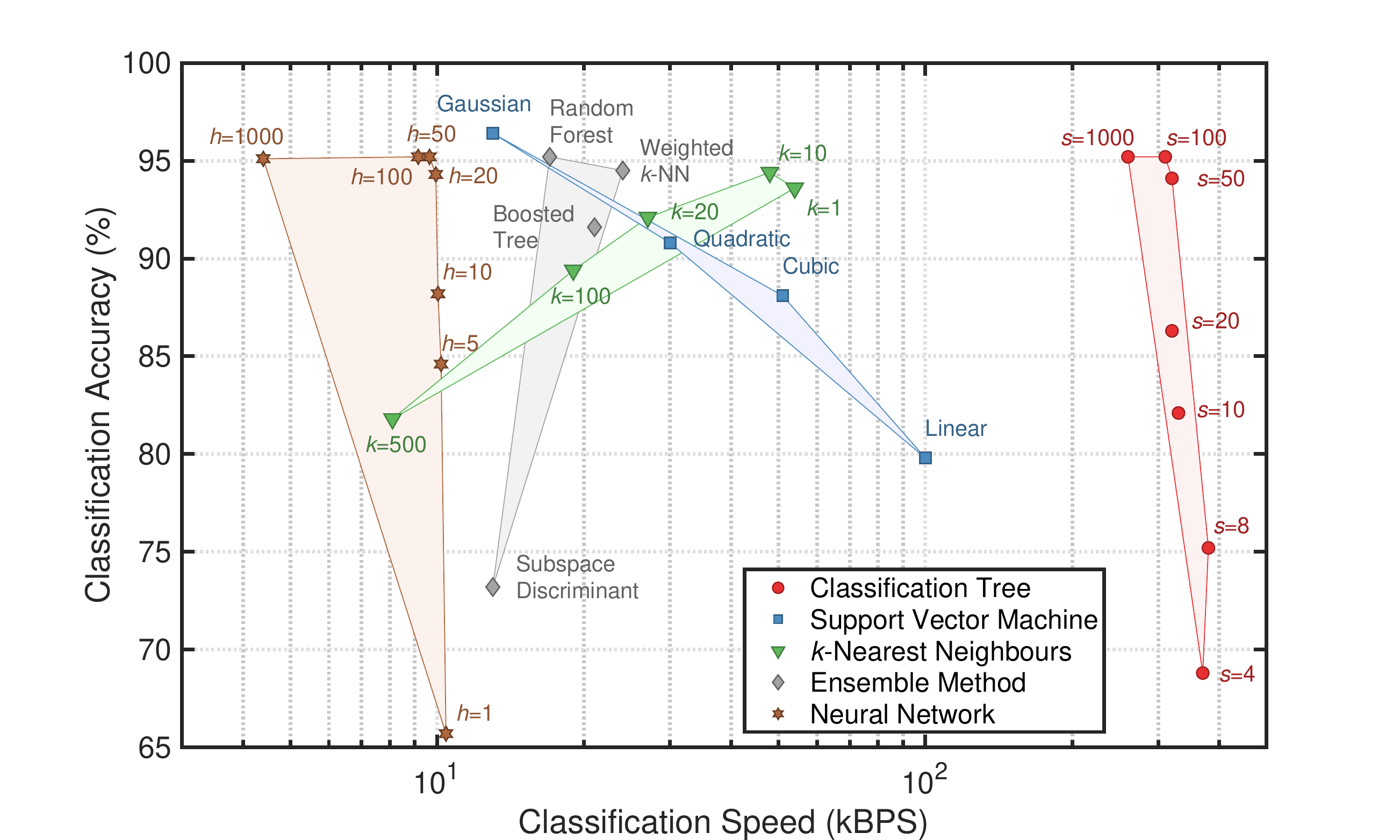}
\caption{An overview of the accuracy and classification speed in kBPS (bursts-per-second) achievable with different supervised-learning methods. The parameter $s$ represents the maximum number of splits in the classification trees and with $h$ the number of hidden layers of the neural networks.}
\label{FIG:complexity}
\end{figure}

The authors employ the dataset of~\cite{Simone_RT_IDI} to show the accuracy/complexity trade-off. 
The dataset includes interference from IEEE 802.15.4, IEEE 802.11b/g/n, and IEEE 802.15.1/BLE sources and it is collected with IEEE 802.15.4 commodity hardware in industrial, semi-industrial and RF-controlled environments. 
It includes over \num{5000} data-points corresponding to interference-bursts represented in a feature-space with \num{8} dimensions, including the previously discussed SFs and envelope-based features.
The dataset is used as for training several machine leaning methods that operate a multi-class burst-specific classification, meaning that the technology of each recorded burst is inferred based on its feature-space representation.

Fig.~\ref{FIG:complexity} shows the performance of the different methods in terms of classification accuracy and average classification speed.
The latter metric, measured offline on a dedicated machine, shows the average time needed to classify one signal burst with an employed ML method.
The figure benchmarks ML methods including neural networks with a different number of hidden layers (parameter $h$ in Fig.~\ref{FIG:complexity}), SVMs with different kernel functions, $k$-nearest neighbours classifiers ($k$-NN) and ensemble methods including random forests of classification trees (CTs), boosted trees, subspace discriminant and weighted $k$-NN.
Such methods are compared with simple CTs of different depth (which scales up with $s$, the maximum number of decision splits) showing a solid baseline performance despite their simple structure.
In this context, there is also a limited advantage of using CT with more than \num{50} split points as the gain on classification accuracy is less than \SI{2}{\percent}.
The reason is that the SFs provide a significant separation in the feature-space, ensuring a good accuracy even with low-complexity classification schemes such as CTs with less than 20 split-points. 
The CTs have been successfully tested on constrained WSN platforms~\cite{Simone_RT_IDI} and are exploited for run-time interference mapping later in this work.

\begin{figure}[!t]
\centering
\includegraphics[width=1\textwidth,clip, trim=0cm 0cm 0cm 0cm]{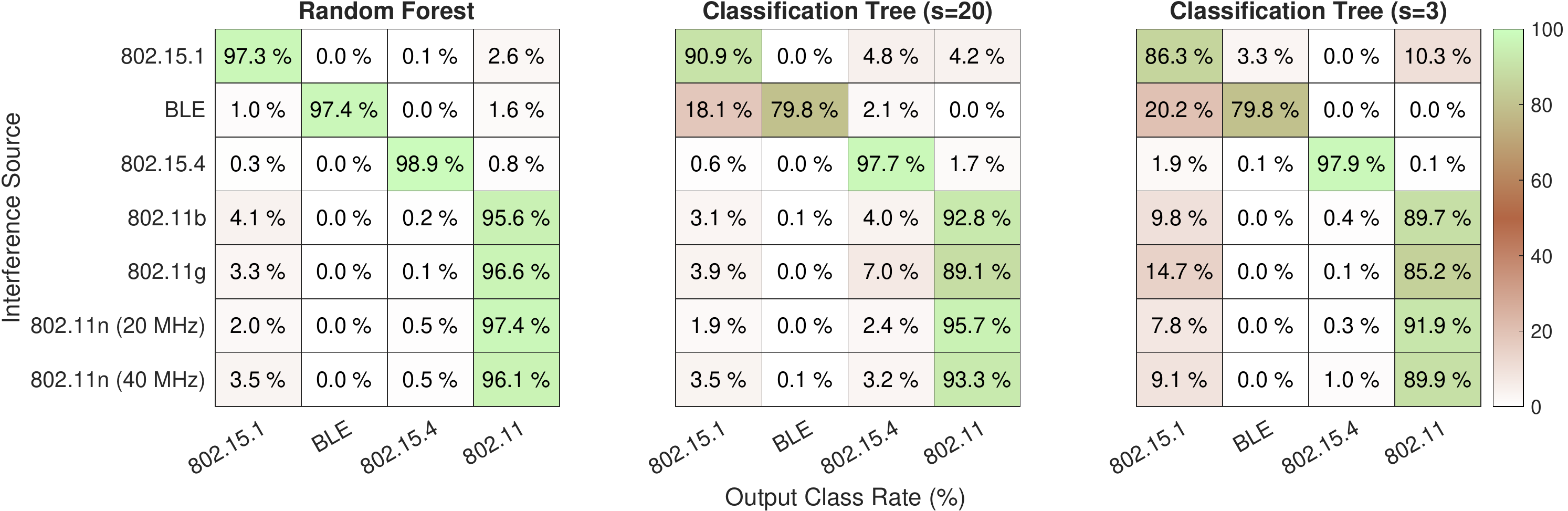}
\caption{Comparison of the interference-specific classification rate between random forest and classification trees with $s$ split points.}
\label{FIG:technlogy_clas}
\end{figure}

Fig.~\ref{FIG:technlogy_clas} clarifies the performance of the candidate classifiers in the presence of different variants of IoT wireless standards.
As shown, more elaborate methods enable improved accuracy in the identification of interference sources of similar feature footprint (e.g., IEEE 802.15.1 and BLE) and lower accuracy deviation among the WLAN amendments (e.g., IEEE 802.11b/g/n).
The analysis is important in the run-time implementation in IIoT, as elaborate ML methods significantly impact the processing-time.
As the classification is operated on a burst-basis, a device performs several classification instances per sensing round. 
When the classification is performed in real-time (as in this work), a limited processing time can also lower power consumption by reducing the time of the spectrum-sensing task, i.e., the onboard radio in active state may drain a significant amount of energy~\cite{energy}. 
Even when the classification is executed offline---the radio bursts are recorded and stored for a posteriori analysis---a short processing time implies a lower stress for the processing unit and higher degree of freedom to schedule and execute standard network tasks (i.e., transmitting/receiving packets). 

\subsection{Real-time Classification}
If the classification time is sufficiently short, burst-based ITI enables real-time identification of the source of interfering transmissions. 
A real-time ITI requires that the processing time needed for burst-classification is comparable to the duration of the interfering transmission from the target technology.
For example, \cite{Simone_RT_IDI} shows that low-complexity IIoT platforms~\cite{telosB} can perform technology classification based on a \num{5} levels CT in a mean processing time of \SI{0.6}{\milli\second}. Considering a duration of identifiable bursts ranging from \SI{0.35}{\milli\second} to \SI{5}{\milli\second}, the method is a real-time ITI.

The real-time on-device processing is vital for the approaches requiring fast detection of interfering transmissions, such as local inference-avoidance. For example, when an IEEE 802.11 packet is detected, the frequency occupancy of the interfering WLAN can be directly inferred, and the transmission channel can be adapted on a link-basis.
On-device processing is also a key enabler for the construction of interference maps as only the classification-output, but not the complete sample-traces needs to be sent to the IIoT network manager, sparing network resources.
Besides, real-time classification enables local estimation of the separate statistics of the interarrival-time of coexisting IoT technologies, which is a key-information for opportunistic transmission schemes.

\section{Opportunities and Challenges of Interference Mapping}
\label{sec:ChallengesAndOppertunities}
\begin{figure}[]
\centering
\includegraphics[width=0.85\textwidth, clip, trim = 0cm 7.8cm 0cm 4.5cm]{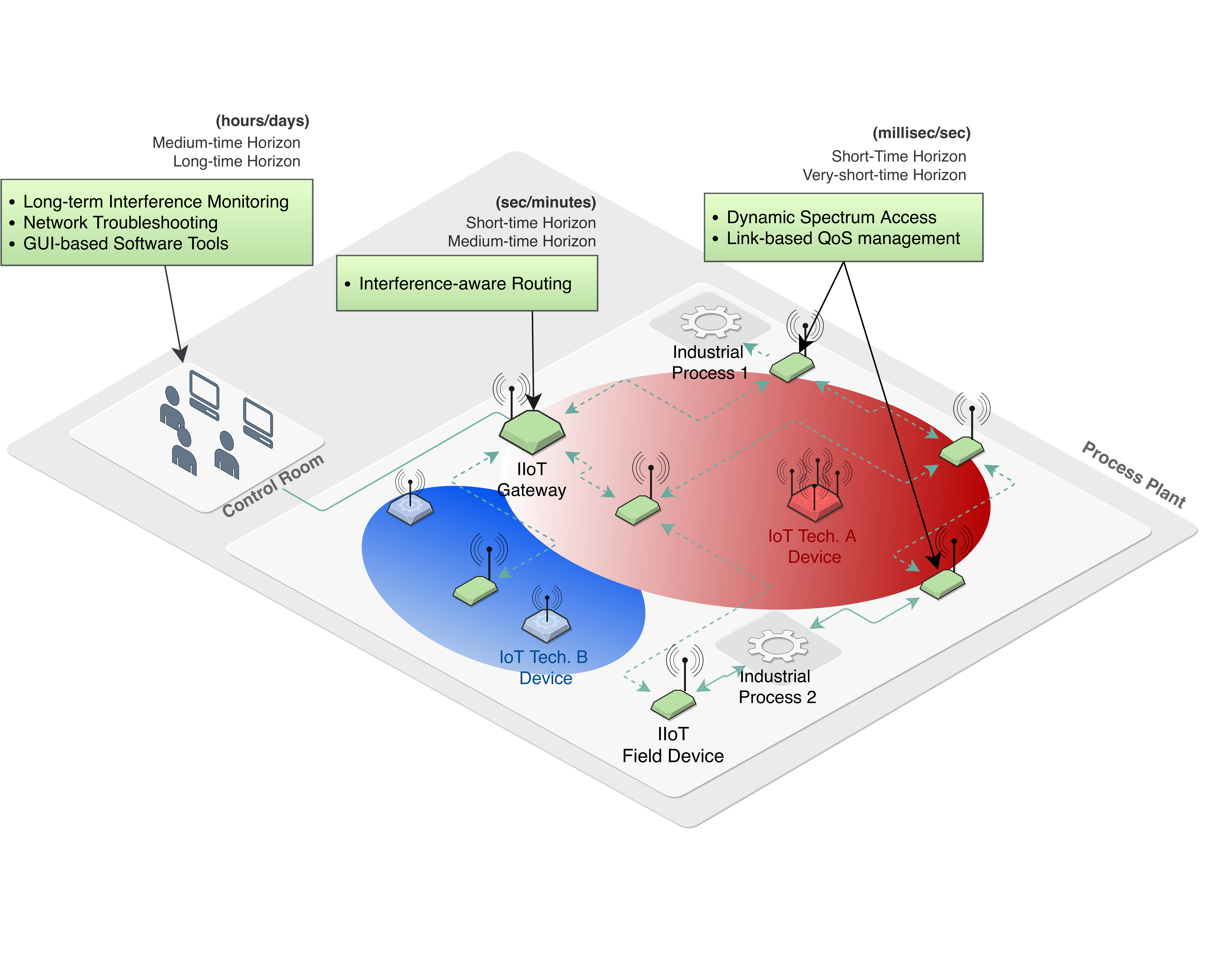}
\caption{Different possibilities of using interference maps in IIoT cognitive radio networks.}
\label{FIG:use_case}
\end{figure}
Since interference maps are complex multidimensional objects, their run-time maintenance can encounter significant obstacles in IIoT networks.
In other words, updating maps with unnecessarily high requirements (e.g., excessive spatial- or time-resolution) can become pointlessly demanding for the network to the point of becoming unfeasible.
Generally, the requirements are defined by the method or application making use of interference maps.
For example, different cognitive approaches require different levels of detail and freshness of the observation data (i.e., the interference map).
With this respect, a categorization of the cognitive radio techniques in the literature~\cite{Cognitive_Industrial} is possible with regard to the target time-horizon of their effect. 
The conventional idea of many approaches is indeed to estimate the current state of the radio resource or predict its future state~\cite{spec_infer}, based on current- and past-observations and to adapt the transmission scheme accordingly. 
Therefore, this work suggests an arbitrary division in three main categories.
The classification is also used as a basis to examine the possible applications of the interference maps in IIoT contexts, which is also sketched in Fig.~\ref{FIG:use_case}. 

\subsubsection{Short- and very-short time-horizon} This class of methods targets the prediction of blank spaces (i.e., channel-idle times) to achieve opportunistic transmission on a very short time-scale.
The approach originates from the assumption that the dynamics of interference sources change rapidly, and the validity of the prediction expires within tens or hundreds of milliseconds.
Because of the tight time-constraint, the decision should be possibly taken on the device level.
Therefore, the construction of global interference maps in such cases appears especially challenging due to the sustained exchange rate of interference reports, particularly in large IIoT mesh networks.

\subsubsection{Medium time-horizon} These methods target dynamic and autonomous allocation of network resources over a time-span of tens of seconds or minutes.
Typical examples of this class are the interference-aware routing solutions, where the optimal allocation of radio channels (e.g., via blacklisting~\cite{blacklist}) is performed on a link basis.
The construction and use of global interference maps are more feasible here since the time-constraints are relaxed.
The approach can potentially improve both the speed and precision of network adaptation compared to the methods based on a posteriori interference assessment (\!\!\cite{ISA100,blacklist}), as in the latter case the network needs to wait for the effects of the interference to manifest before addressing them.
Besides, the dedicated network resources need to be allocated efficiently for aggregating interference reports under the coexistence with regular traffic.

\subsubsection{Long time-horizon} The last class of the methods concerns the mapping of long-term (e.g., day/night time) variations of channel load in space and frequency. The time constraints are here relatively minor. Yet, the separation on the interference technology plane provides a powerful tool for manual IIoT network planning and troubleshooting of coexistence issues. Notably, the solution ensures a tunable resolution in the time-domain while at the same time avoids the hassle of deploying additional measurement hardware.


\section{Interference Mapping with ML-ITI}
\label{sec:InterferenceMaps}

The authors have tested a ML-ITI solution in the construction of interference maps with an IEEE 802.15.4 star-network deployed indoor. Particularly, an ITI-based mapping solution in analysis has been implemented in 15 battery-powered Crossbow TelosB~\cite{telosB}---low-cost IIoT devices. 
An overview of the platform's characteristics and sensing parameters is reported in the rightmost column of Table~\ref{TAB:parameters}.
The devices are deployed in an office setting, covering a total area of \SI{12x25}{\meter} composed of corridors, several offices, and a coffee room. The network is configured with a star topology, where the central device operates as a sink/gateway, while the 14 sensor nodes are deployed in offices and corridors.
The deployed network coexists with wireless systems of two types. First, three IEEE 802.11-based access points are in place to provide Internet connectivity to the personnel, and both stationary and mobile WLAN devices are present in the area.
Second, an IEEE 802.15.1-based point-to-point network ensures wireless data transfer between two portable devices.

\subsection{Spectrum Sensing and Classification}
The employed real-time ITI method based on classification trees is used to secure a non-invasive sensing process. In particular, the network devices perform a fast spectrum scan, sequentially covering all the \num{16}-IEEE 802.15.4 channels with an observation time of \SI{50}{\milli\second} per channel. During the sensing process, the ITI enables a real-time assessment of the power, duration, and source-technology of each detected signal burst. Subsequently, the device updates a local technology-frequency-time interference matrix, transmits a data-packet containing an interference report to the IIoT gateway, and switches to idle-mode.
Since the sensing process has a \SI{5}{\second} periodicity, the duty-cycle of the whole sensing, identification, and reporting process is moderate (around \SI{15}{\percent}).
The choices of duty-cycle and sensing periodicity are related to the desired time-resolution of the interference maps, as well as the network requirements, as discussed in Section~\ref{sec:ChallengesAndOppertunities}. 

Although all the interference bursts detected during the sensing process are classified and timestamped, the data transmitted with each report is kept to a minimum. In particular, a time-average over the sensing interval is carried out, while retaining information on interference source and frequency distribution. The choice is motivated by both the packet-size limit of the IEEE 802.15.4 standard and by the requirement of constructing interference maps with moderately detailed information in frequency and space-domain.
In other words, in the tested use case, the short-term evolution of the interference is less relevant than its long-term space-frequency distribution.

\subsection{Map Construction}
\begin{figure*}[ht]
  \centering
       \includegraphics[width=0.9\linewidth,clip, trim=0cm 0cm 0cm 0cm]{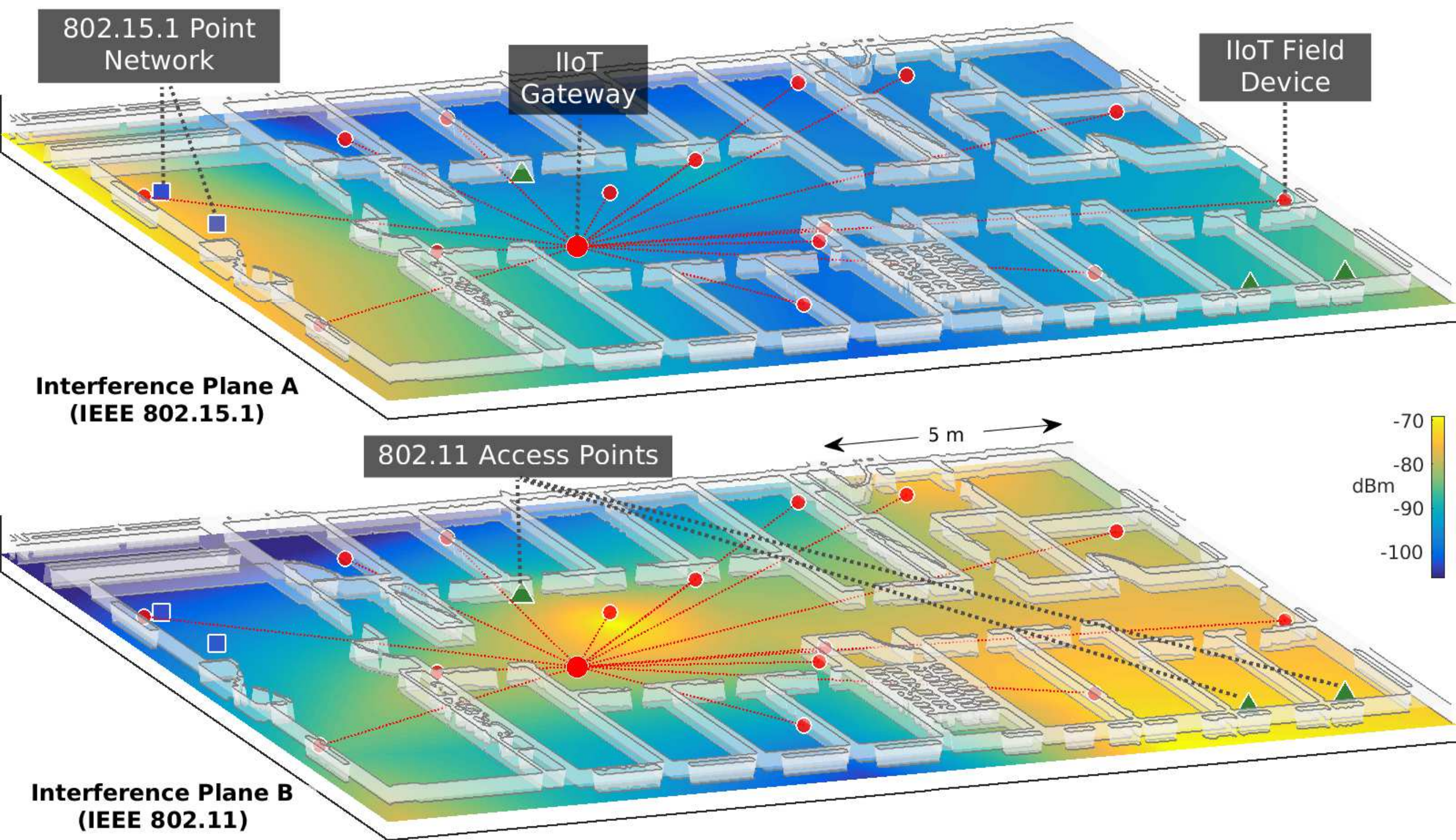}
        \caption{Power maps of coexisting interference technologies (IEEE 802.11 and IEEE 802.15.1) classified in real-time. The position of the devices running the ITI mapping algorithm is marked in red.}
        \label{FIG:interference_map_power}
\end{figure*}

\begin{figure}[ht]
  \centering
        \includegraphics[width=0.9\linewidth,clip, trim=2cm 0cm 0.6cm 0.5cm]{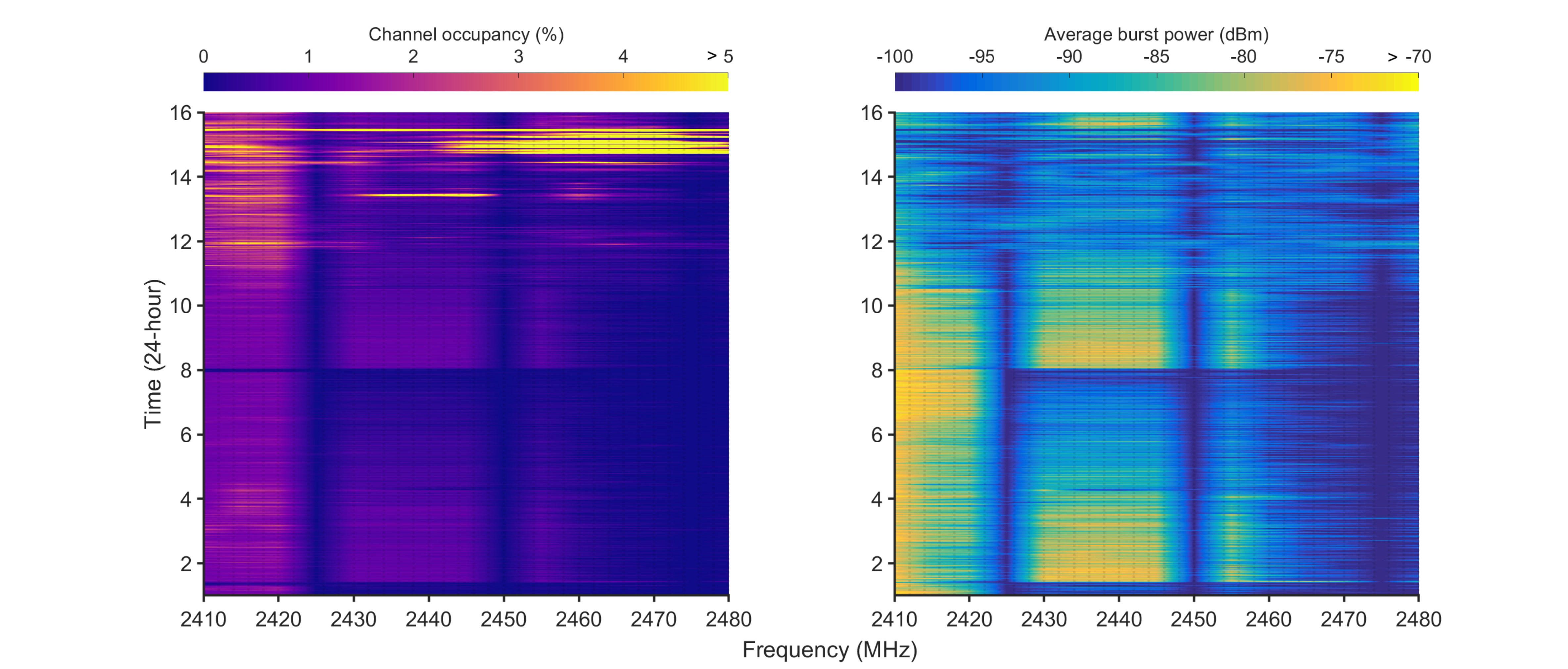}\label{FIG:frequency_time_map_wifi}
  \caption{Spectrograms of power and channel-occupancy (busy-time) of IEEE 802.11 WLANs in one point of the environment, as reported by one of the IIoT field devices.}
  \label{FIG:frequency_time_map}
\end{figure}
Basing on interference reports, the IIoT network-manager generates and updates multidimensional matrices containing information in frequency-, node ID-, technology-, and time-domain.
There are, in general, several possibilities for combining the information of the interference reports. Since maintaining multidimensional matrices at run-time requires computational efforts from the network manager, a recommended practice is to tailor the structure of such matrices to the specific cognitive radio use case.
In this example, two \num{4}-dimensional matrices (time, frequency, IoT technology, node ID) are generated from the reports. The two matrices represent the average power of interference bursts and the channel busy-time, respectively. 
The \textit{a priori} known position of the sensing nodes together with data-interpolation allows constructing a continuous map of the average power of interference bursts in the space-domain. The interpolation method of choice is the natural-neighbor~\cite{nearest_nei}, which is reported to have optimal performance for irregularly distributed sample points.
Fig.~\ref{FIG:interference_map_power} shows such a power map, collected in the interference scenario outlined above. 
The map is averaged over both the \si{16}-frequency channels and over a \num{15} minutes observation window, providing an example of a human-readable representation of complex multidimensional data.

Fig.~\ref{FIG:frequency_time_map} focuses on the analysis of a larger time-window and shows a partial snapshot from around \SI{42}{\hour} of continuous network run-time. The spectrograms reflect the content of both the average power and the channel busy-time matrices on the IEEE 802.11 interference dimension. 
The spectrograms show how the average power alone provides a merely partial picture of the dynamics of interfering WLANs, since the channel-busy time and emitted power can evolve independently. The reasons are the sporadicity of WLAN traffic and the mobility of most WLAN devices (e.g., smartphones).

The assessment of the technology-specific busy/idle-times of the interfering network is particularly relevant to predict the performance impact on the victim-IIoT network. Indeed, several related works~\cite{WLAN_vs_WSN,mutual_int} show that the variation of idle-time of IEEE 802.11-based WLANs has a significant effect on the caused performance reduction; the effect is instead of more limited entity when the interfering technology is IEEE 802.15.1.

\section{Conclusion}
\label{sec:conclusion}
This work analyzed the relevance, feasibility, and potential uses of real-time wireless technology classification for RF interference mapping in IIoT networks over unlicensed bands.
The paper discussed how the on-device information on the origin of interference and its time-frequency characteristics can enable dynamic spectrum access, interference-aware routing, or enhanced network coexistence on different layers of the protocol stack.
Furthermore, the benefits and feasibility of machine learning on the on-device classification is analyzed, and a real use case comprising IEEE 802.15.4 battery-powered devices is presented. 
The experiments have shown that even extremely resource-constrained devices can recognize the nature of cross-technology interference at the level of signal bursts at run-time.
The information is used to perform advanced radio-environment analysis by extracting information such as technology-specific inter-arrival time and spectrum occupancy in real-time, allowing the fine-grained tracking of interference in time, frequency, and space.
The results constitute a promising step in the prospect of massive IoT deployment and QoS-enforcement in industrial IoT applications in radio environments with many coexisting wireless technologies.

\bibliographystyle{IEEEtran}
\bibliography{MagBib}

\end{document}